        \newcommand{\DOT}{\hspace{-0.08in}{\bf .}\hspace{0.1in}}
 	\newcommand{\BOX}{\hbox {$\sqcap$ \kern -1em $\sqcup$}}
	\newcommand{\qed}{\hskip 3em \hbox{\BOX} \vskip 2ex}
	
	\newcommand\C{{\bf C}}
	\def\section#1{\vskip3em{\centerline {\bf#1}}\vskip3em}
	
	\newcommand{\be}{\begin{equation}}
        \newcommand{\ee}{\end{equation}}
        \newcommand{\ba}{\begin{eqnarray}}
        \newcommand{\ea}{\end{eqnarray}}
        \newcommand{\ban}{\begin{eqnarray*}}
        \newcommand{\ean}{\end{eqnarray*}}
	\newcommand{\A}{{\cal A}}
	\newcommand{\G}{{\cal G}}
        \newcommand{\maps}{\colon}

	\newcommand{\Diff}{{\rm Diff}}
	
	\newcommand{\Aut}{{\rm Aut}}
	\newcommand{\Fun}{{\rm Fun}}
	\newcommand{\Ad}{{\rm Ad}}
	
	\newcommand{\D}{{\cal D}}
	\newcommand{\om}{\omega}
	\newtheorem{theorem}{Theorem}
	
	\newtheorem{lemma}{Lemma}

	\documentstyle[12pt]{article}
\begin{document}

	\begin{center}
	{\bf  Generalized Measures in Gauge Theory \\}
	\vspace{0.5cm}
	{\em John C. Baez\\}
	\vspace{0.3cm}
	{\small Department of Mathematics \\
	University of California\\
        Riverside CA 92521\\}
	\vspace{0.3cm}
	{\small October 30, 1993 \\}
	\vspace{0.3cm}
	\end{center}

\begin{abstract}
Let $P \to M$ be a principal $G$-bundle.  We construct
well-defined substitutes for ``Lebesgue measure'' on the space
$\A$ of connections on $P$ and for ``Haar measure'' on the group $\G$ of
gauge transformations.  More precisely, we define algebras of ``cylinder
functions'' on the spaces $\A$, $\G$, and $\A/\G$,
and define generalized measures on these spaces as continuous linear
functionals on the corresponding algebras.
Borrowing some ideas from
lattice gauge theory, we characterize generalized measures on $\A$,
$\G$, and $\A/\G$ in terms of graphs embedded in $M$.  We use this
characterization to construct generalized measures
on $\A$ and $\G$, respectively.  The ``uniform'' generalized measure
on $\A$ is invariant under the group of automorphisms of $P$.  It
projects down to the generalized measure on $\A/\G$ considered by
Ashtekar and Lewandowski in the case $G = SU(n)$.  The
``generalized Haar measure'' on $\G$ is right- and left-invariant as
well as $\Aut(P)$-invariant.  We show that averaging any generalized
measure on $\A$ against generalized Haar measure
 gives a $\G$-invariant generalized measure on $\A$.
\end{abstract}

\section{Introduction}

The space $\A$ of connections on a principal bundle is an
infinite-dimensional affine space, and the while the notion of the
``Lebesgue measure'' $\D A$ on $\A$ has been very fruitful, it is
mathematically ill-defined.  Some of the infinities in quantum field
theory calculations can be avoided by projecting down $\D A$ to the
space $\A/\G$ of connections modulo gauge transformations, but certainly
not all.  While the theory of cylinder measures on infinite-dimensional
vector spaces \cite{Kolm} provides a rigorous framework for interpreting
the Gaussian ``measures'' appearing in the physics of the free boson
field \cite{BSZ}, it is usually quite difficult to apply this theory to
the study of gauge fields, except in the case of 2d Yang-Mills theory.
It is thus desirable to generalize the concept of measure in a manner
more suited to the needs of gauge theory.

Recently Ashtekar and Isham \cite{AI} have proposed an approach based on
the idea of a Wilson loop, that is, the trace of the holonomy of a
connection around a loop in the base manifold $M$.  Wilson loops are
very natural observables in gauge theory, and in the ``loop
representation'' of gauge theories they play a primary role \cite{Loll}.
In the case of the group $G = SU(n)$, Ashtekar and Lewandowski \cite{AL}
used this approach to define and construct a very natural ``generalized
measure'' $\mu_{\rm AL}$ on $\A/\G$, which is invariant under
diffeomorphisms of $M$.  The author \cite{B} extended this approach to
construct a rich variety of diffeomorphism-invariant generalized
measures on $\A/\G$ when $G$ is compact.  These generalized measures
give rise to invariants of multiloops (collections of loops) in the base
manifold, and their classification involves a combination of singularity
theory and knot theory.

In this paper we show that one can define generalized measures on $\A$.
All of these project down to generalized measures on $\A/\G$, but even
when one is interested in gauge-invariant quantities, it is sometimes
easier to work ``upstairs'' on $\A$.  In particular, when $G$ is
compact, there is a ``uniform'' generalized measure $\mu_u$ on $\A$ that
projects down to $\mu_{AL}$ under the map $\A \to \A/\G$.  This
generalized measure $\mu_u$ is in some respects a rigorous substitute
for the ill-defined ``Lebesgue measure'' on $\A$, but it is actually
built using Haar measure on $G$.  We also define generalized measures on
$\G$, and when $G$ is compact we construct a natural example $\mu_{\rm
H}$ that is a rigorous substitute for Haar measure on $\G$.  As an
application of this ``generalized Haar measure'' we show that any
generalized measure on $\A$ can be averaged against $\mu_{\rm H}$ to
give a $\G$-invariant generalized measure on $\A$.

We emphasize that these constructions are no panacea; in particular,
they are unlikely to be of much use in 4-dimensional Yang-Mills theory,
where one expects that only ``smeared'' Wilson loops will serve as
physical observables \cite{B2}.  With some modification, these
constructions might allow the construction of the Chern-Simons path
integral as a generalized measure, as discussed in \cite{B}.  They may
also be suited for rigorous work on the loop representation of quantum
gravity \cite{Loll,RS}.

\section{Generalized Measures} \label{prelim}

Let $M$ be a manifold, possibly with boundary, let $G$ be a Lie group,
and let
$\pi \maps P \to M$ be a principal $G$-bundle.
Let $\A$ be the space of connections on $P$ and let $\G$ be the
group of gauge transformations.  Let $\Diff(M)$ denote the group
of diffeomorphisms of $M$ restricting to diffeomorphisms
of $\partial M$.   Everything in this section applies
equally to the following three cases:
\begin{enumerate}
\item The $C^\infty$ case: $M$ and $P$ are smooth, $\pi$ is smooth,
and $\A$, $\G$, and $\Diff(M)$ consist of smooth connections, gauge
transformations, and diffeomorphisms, respectively.

\item The $C^\omega$ case: $M$ and $P$ are real-analytic, $\pi$ is
real-analytic, and $\A$, $\G$, and $\Diff(M)$ consist of real-analytic
connections, gauge transformations, and diffeomorphisms, respectively.

\item The hybrid case: $M$ is real-analytic and $P$ is smooth, $\pi$ is
smooth, $\A$ and $\G$ consist of smooth connections and gauge
transformations, respectively, and $\Diff(M)$ consists of real-analytic
diffeomorphisms.
\end{enumerate}

We will define sub-C*-algebras of the bounded continuous complex
functions on $\A$, $\G$, and $\A/\G$ (with their $C^\infty$ topologies).
By a ``generalized measure'' on one of these spaces we will mean simply
a continuous linear functional on the corresponding C*-algebra.  Every
finite regular Borel measure on one of these spaces defines such a
generalized measure, but the most
interesting generalized measures are not of this form.  It is easiest to
construct generalized measures in case 3 above, so in the next section
we will restrict attention to that case, even though case 1 is in some
ways the most natural.

Given a path $\gamma \maps [0,1] \to M$, let $\A_\gamma$ be the
space of all maps from the fiber $P_{\gamma(0)}$ to the fiber
$P_{\gamma(1)}$ that can be obtained as holonomies along $\gamma$ of
some connection $A \in \A$.   Note that there is a
natural map
\ban       p_\gamma \maps \A &\to& \A_\gamma   \\
                     A& \mapsto& A_\gamma  \ean
assigning to each connection $A \in \A$ its holonomy $A_\gamma$
along $\gamma$.
One should think of $p_\gamma\maps A \mapsto A_\gamma$ as picking
out a small piece of
information about the connection $A$; one can reconstruct $A$ from all
the pieces $\{A_\gamma\}$.  Fixing trivializations of $P$ at the
endpoints $\gamma(0)$ and $\gamma(1)$, $\A_\gamma$
can be identified with an open and closed
subspace of the group $G$.  We give $\A_\gamma$
the subspace topology (which is independent of the choice of
trivialization).  This makes the map $p_\gamma$
continuous.

We say $f$ is a {\it cylinder} function on $\A$ if it is of the form
\[        f(A) =   F(A_{\gamma_1}, \dots , A_{\gamma_n}), \]
where $\{\gamma_i\}$
is a finite set of paths in $M$ (required to be real-analytic in the
$C^\om$ and hybrid cases) and
\[    F \maps \prod_i \A_{\gamma_i}  \to \C \]
is a bounded continuous function.
Let $\Fun_0(\A)$ denote the space of cylinder functions
on $\A$, which is
a $\ast$-subalgebra of the bounded continuous functions on $\A$.
The completion of $\Fun_0(\A)$ with
respect to the $\sup$ norm, which we denote by $\Fun(\A)$, is thus a
C*-subalgebra of the bounded continuous functions on $\A$.

Note that $\G$ acts as $\ast$-automorphisms of $\Fun_0(\A)$ by
\[                gf(A) = f(g^{-1}A) .\]
The $\G$-invariant functions in $\Fun_0(\A)$ may be regarded as
functions on $\A/\G$, and we denote the algebra of all such functions as
$\Fun_0(\A/\G)$.  We call these {\it cylinder} functions on $\A/\G$.
We denote the
 completion of $\Fun_0(\A/\G)$ with respect to the $\sup$ norm by
$\Fun(\A/\G)$.  This may be regarded either as
 a C*-subalgebra of the bounded continuous functions on $\A/\G$ (with
its quotient topology), or of
the bounded continuous $\G$-invariant functions on $\A$.
It can be seen that as
special cases of this algebra one
obtains the ``holonomy C*-algebra'' defined in the smooth case
by Ashtekar and Isham \cite{AI} for $G = SU(2)$ and the ``analytic
holonomy C*-algebra'' defined in the hybrid case
by Ashtekar and Lewandowski \cite{AL} for $G = SU(n)$.

We may also define cylinder functions on the
group $\G$ of gauge transformations.
Given a point $x \in M$, let $\G_x$ be the fiber at $x$ of
the bundle $P \times_{\Ad} G$, with its subspace topology.
An element $g \in \G$
is a section of $P \times_{\Ad} G$, so there is a natural map
\ban       p_x \maps \G &\to& \G_x   \\
                     g& \mapsto& g_x  \ean
One should think of $p_x \maps g \mapsto g_x$ as picking out a small
piece of information about the gauge transformation $g$; one can
reconstruct $g$ from the pieces $\{g_x\}$.
Note that $\G_x$ is naturally a group, and that $p_x$
is a homomorphism.

We say $f$ is a {\it cylinder} function on $\G$
if it is of the form
\[      f(g) =   F(g_{x_1}, \dots, g_{x_n}) ,\]
where $\{x_i\}$ is a finite set of points in $M$ and
\[    F \maps \prod_i \G_{\gamma_i}  \to \C \]
is bounded and continuous.
The completion of the algebra $\Fun_0(\G)$ of
cylinder functions on $\G$ is a C*-subalgebra of the bounded continuous
functions on $\G$, which we denote by $\Fun(\G)$.

By a {\it generalized measure} on $\A$, $\G$, or $\A/\G$ we mean a
continuous linear functional on $\Fun(\A)$, $\Fun(\G)$, or
$\Fun(\A/\G)$, respectively.  Note that every generalized
measure on $\A$ ``projects down'' to a generalized measure on $\A/\G$;
this operation of ``projecting down'' is really just restriction of a
continuous linear functional on $\Fun(\A)$ to the
subalgebra $\Fun(\A/\G)$.

Let $\Aut(P)$ denote the group
of bundle automorphisms $g$ such that for some $h \in \Diff(M)$,
$\pi(g(p)) = h(\pi(p))$ for all $p \in P$.  (In cases 2 and 3, recall
that $h$ must be real-analytic.)  Then we have an exact
sequence
\[      1 \to \G \to \Aut(P) \to \Diff(M) \to 1 .\]
The group
$\Aut(P)$ acts as $\ast$-automorphisms of $\Fun(\A)$ by
\[                gf(A) = f(g^{-1}A) .\]
so we obtain an action
of $\Diff(M)$ as $\ast$-automorphisms of $\Fun(\A/\G)$.
By duality, $\Aut(P)$ acts on the generalized measures on $\A$, and
$\Diff(M)$ acts on the generalized measures on $\A/\G$.
Any $\Aut(P)$-invariant generalized measure on $\A$ projects down to a
$\Diff(M)$-invariant generalized measure on $\A/\G$.
Note also that $\Aut(P)$ acts on the generalized measures on $\G$, as do
left and right translation.

\section{Characterizing Generalized Measures}

In this section and the next we restrict our attention to the
``hybrid case,'' case 2 of the previous section.  Ashtekar and
Lewandowski recognized the importance of this case when they used it to
construct a very natural sort of $\Diff(M)$-invariant generalized
measure on $\A/\G$ for $G = SU(2)$.  Subsequently they, and
independently the author, were able to generalize this construction to
more general compact Lie groups, and also to give a rather concrete
characterization of all generalized measures on $\A/\G$.
In addition, the author has given a recipe for constructing many
$\Diff(M)$-invariant examples of such generalized measures.

Here we give concrete characterizations of generalized measures on
$\A$, $\G$, and $\A/\G$ when $G$ is any Lie group.
First, we need a notion of an embedded graph in $M$, a slight
variant of that in \cite{B}.  We define an
{\it embedded graph} $\phi$ in $M$ to be a finite
collection of real-analytic paths $\phi_j \maps [0,1] \to M$ such that:
\begin{enumerate}
\item  for all $j$,  $\phi_j$ is one-to-one,
\item for all $j$, $\phi_j|_{(0,1)}$ is an embedding,
\item for all $j$ and $k$,
$\phi_j[0,1] \cap \phi_k[0,1] \subseteq \{\phi_j(0),\phi_j(1)\}$.
\end{enumerate}
\noindent The paths $\phi_j$ are called the {\it edges} of $\phi$,
and the points $\phi_j(0),\phi_j(1)$ are called the {\it vertices}
of $\phi$.  Somewhat redundantly, we
write $E(\phi)$ for the set of edges of $\phi$ and
$V(\phi)$ for the set of (distinct) vertices.  Note that the set
\[    |\phi| =  \bigcup_j \phi_j[0,1] \subseteq M \]
equipped with the subspace topology indeed has the topology of a finite
graph.

The following lemma proved by Ashtekar and Lewandowsi plays a
key technical role.

\begin{lemma} \label{lem1} \DOT
{\rm \cite{AL}} Let $\{\gamma_i\}$ be a finite
collection of real-analytic paths in $M$.  Then there exists an embedded
graph $\phi$ such that for each $\gamma_i$ there exist paths in $\phi$
such that $\gamma_i$ is equivalent to a product of these paths and their
inverses, up to a continuous orientation-preserving reparametrization.
\end{lemma}

Given an analytic graph $\phi$ in $M$, let
\[     \A_\phi = \prod_{\gamma \in E(\phi)} \A_{\gamma} ,\]
the Cartesian product over all edges $\gamma$ of $\phi$ of the spaces
$\A_\gamma$, equipped with the product topology.  Similarly, let
\[      \G_\phi = \prod_{x \in V(\phi)}  \G_x \]
equipped with the product topology.
We may write any element of $\G_\phi$ as a tuple
$(g_x)_{x \in V(\phi)}$ where $g_x \in \G_x$.  Similarly, we may write any
element of $\A_\phi$ as a tuple $(A_\gamma)_{\gamma \in E(\phi)}$, where
$A_\gamma \maps P_{\gamma(0)} \to P_{\gamma(1)}$.
There are natural maps $p_\phi \maps \A \to \A_\phi$ and
$p_\phi \maps \G \to \G_\phi$, given by
\ban         p_\phi(A) &=& (A_\gamma)_{\gamma \in E(\phi)}  ,\\
         p_\phi(g) &=& (g_x)_{x \in V(\phi)}  .\ean
Though we denote both of these maps by $p_\phi$, the meaning
should be clear from context.
Both these maps are onto, since we
can always find a connection having any specified holonomies in
the sets $\A_\phi$, and we can always find a gauge transformation having any
specified values at the vertices of $\phi$.  Given $A \in \A$ and $g \in
\G$, we will sometimes write
$A_\phi$ for $p_\phi(A)$ and $g_\phi$ for $p_\phi(g)$.

In the above we are borrowing an idea from lattice gauge theory, in
which ``connections'' assign group elements to the edges of a lattice,
while ``gauge transformations'' assign group elements to vertices.  We
can make this analogy very precise if the group $G$ is connected.
In this cases, we can trivialize $P$ over
$|\phi|$ for any embedded graph $\phi$.  Fixing a trivialization gives
an identification of $\A_\gamma$, for any edge $\gamma$ of $\phi$, with
the group $G$, hence
\[        \A_\phi \cong G^{E(\phi)} .\]
Similarly, fixing a trivialization gives an identification of
$\G_x$, for any vertex $x$ of $\phi$, with $G$, so
\[       \G_\phi \cong G^{V(\phi)} .\]

The group $\G_\phi$
acts on the space $\A_\phi$ as follows:
\[       (g_x)_{x \in V(\phi)} \,(A_\gamma)_{\gamma \in E(\phi)}
 =  (g_{\gamma(1)}  A_\gamma g_{\gamma(0)}^{-1})_{\gamma \in E(\phi)}  .\]
This action is compatible with the action of $\G$ on $\A$, as follows:
\[         p_\phi(g) p_\phi(A) = p_\phi(gA) \]
for any $g \in \G$, $A \in \A$.

Let $\Fun(\A_\phi)$ denote the algebra of
bounded continuous functions on $\A_\phi$.  We will identify
$F \in \Fun(\A_\phi)$ with the function $f$ on $\A$ given by
\[   f(A) = F(A_\phi), \]
allowing us to write
\[      \Fun(\A_\phi) \subseteq \Fun_0(\A)  .\]
Since a generalized measure $\mu$ on $\A$ is just a continuous linear
functional on $\Fun(\A)$, we can restrict $\mu$ to a continuous linear
functional $\mu_\phi$ on $\Fun(\A_\phi)$.  (When $G$ is compact,
$\A_\phi$ is
compact, so the Riesz-Markov theorem allows us to identify continuous
linear functionals on $\Fun(\A_\phi)$ with
finite regular Borel measures on $\A_\phi$.)  A set of
continuous linear functionals $\mu_\phi \in \Fun(\A_\phi)^\ast$, one for
each embedded graph $\phi$, will be called a {\it family}.
The following theorem gives necessary and
sufficient conditions for a family $\{\mu_\phi\}$ to come from a
generalized measure on $\A$.  As in \cite{B}, these conditions
can be used to construct concrete examples of generalized measures.

Given embedded graphs $\phi,\psi$, we say that $\phi$ is {\it included
in} $\psi$, which we write as $\phi \hookrightarrow \psi$, if every edge
of $\phi$ is, up to orientation-preserving reparametrization, a product
of edges of $\psi$ and their inverses.  Note that $\phi \hookrightarrow
\psi$ implies that
every vertex of $\phi$ is a vertex of $\psi$, and that $|\phi|
\subseteq |\psi|$.  It also implies that
$\Fun(\A_\phi) \subseteq \Fun(\A_\psi)$.  We say that the family
$\{\mu_\phi\}$ is
{\it consistent} if $\phi \hookrightarrow \psi$ implies that the
restriction of $\mu_\psi$ to $\Fun(\A_\phi)$ is $\mu_\phi$.
We say that the family $\{\mu_\phi\}$ is {\it uniformly bounded}
if there is a constant $C > 0$
such that $\|\mu_\phi\| < C$ for all $\phi$.

\begin{theorem} \label{thm1} \DOT
Suppose $\mu$ is a generalized measure on $\A$, that is, a continuous
linear functional on $\Fun(\A)$.  For any embedded graph $\phi$ in $M$,
let $\mu_\phi$ denote the restriction of $\mu$ to $\Fun(\A_\phi)$.  Then
$\{\mu_\phi\}$ is a consistent and uniformly bounded family.
Conversely, if $\{\mu_\phi\}$ is a consistent and uniformly bounded
family, there is a unique generalized measure $\mu$ on $\A$ whose
restriction to $\Fun(\A_\phi)$ is $\mu_\phi$.
\end{theorem}

Proof - If $\mu$ is a generalized measure on $\A$ the family
$\{\mu_\phi\}$ obtained by restriction is consistent, and $\|\mu_\phi\|
\le \|\mu\|$, so it is uniformly bounded.  Conversely, suppose we are
given a consistent and uniformly bounded family $\{\mu_\phi\}$.  We
first define a linear functional $\mu$ on $\Fun_0(\A)$ as follows.  Any
element $f \in \Fun_0(\A)$ is of the form
\[          f(A) = F(A_{\gamma_1}, \dots, A_{\gamma_n}), \]
where $\{\gamma_i\}$ are paths in $M$.  In this situation we say that $f$
can be {\it expressed} in terms of the paths $\{\gamma_i\}$.  Construct
an embedded graph $\phi$ from the paths $\gamma_i$ as in Lemma
\ref{lem1}.  Then $f \in \Fun(\A_\phi)$.  Define
\[              \mu(f) = \mu_\phi(f).  \]
We need to check that $\mu$ is well-defined, linear, and extends to a
continuous linear functional on $\Fun(\A)$.  If the extension exists, it
is unique, since $\Fun_0(A)$ is dense in $\Fun(\A)$.

For well-definedness, suppose that $f$ can be expressed in two ways,
in terms of paths $\{\gamma_i\}$ or in terms of paths $\{\gamma'_j\}$.
Using Lemma \ref{lem1}, construct embedded graphs $\phi$ from the paths
$\{\gamma_i\}$, $\phi'$ from the paths
$\{\gamma'_j\}$, and $\psi$ from the paths $\{\gamma_i, \gamma'_j\}$.
Note that $\phi \hookrightarrow \psi$ and $\phi' \hookrightarrow \psi$.
Thus
\[     \mu_\phi(f) = \mu_\psi(f) = \mu_{\phi'}(f) .\]

For linearity, suppose $f,g \in \Fun_0(\A)$.  Then $f+g \in \Fun_0(\A)$
and there exist paths $\{\gamma_i\}$ in terms of which $f,g,$ and $f+g$ can
all be expressed.  Using Lemma 1, construct an embedded graph $\phi$ from
the paths $\{\gamma_i\}$.  Then
\[         \mu(f+g) = \mu_\phi(f+g) = \mu_\phi(f) + \mu_\phi(g) =
\mu(f) + \mu(g) .\]
Clearly $\mu(\lambda f) = \lambda\mu(f)$ for all $\lambda \in \C$.

Finally, to show that $\mu$ extends to a continuous linear functional on
$\Fun(\A)$ it suffices to note that there exists $C > 0$ with
$\|\mu(f)\| \le C\|f\|$ for any $f \in \Fun_0(\A)$, by the uniform
boundedness of the family $\{\mu_\phi\}$.  \qed

Completely analogous results holds for generalized measures on $\G$ and
$\A/\G$.  Let $\Fun(\A_\phi/G_\phi)$ denote the subalgebra of
$\Fun(\A_\phi)$ consisting of functions invariant under the action of
$\G_\phi$.  Alternatively, $\Fun(\A_\phi/\G_\phi)$ may be regarded as
the algebra of bounded continuous functions on $\A_\phi/\G_\phi$,
equipped with its quotient topology.
A generalized measure $\mu$ on $\A/\G$
restricts to a family of elements $\mu_\phi \in
\Fun(\A_\phi/\G_\phi)^\ast$, one for each embedded graph $\phi$.
(When $G$ is compact these are the same as finite regular Borel measures
on $\A_\phi/\G_\phi$.)    We define consistency and uniform
boundedness of such families as before, and obtain:

\begin{theorem} \label{thm2} \DOT
Suppose $\mu$ is a generalized measure on
$\A/\G$.  For any embedded graph $\phi$ in $M$, let $\mu_\phi$ denote
the restriction of $\mu$ to $\Fun(\A_\phi/\G_\phi)$.  Then
$\{\mu_\phi\}$ is a consistent and uniformly bounded family.
Conversely, if $\{\mu_\phi\}$ is  a consistent and uniformly bounded
family, there is a unique generalized measure $\mu$ on $\A/\G$ whose
restriction to $\Fun(\A_\phi/\G_\phi)$ is $\mu_\phi$.
\end{theorem}

Proof - The proof follows that of Theorem \ref{thm1}.
 \qed

Let $\Fun(\G_\phi)$ denote the algebra of bounded continuous functions
on $\G_\phi$.   We will identify $F \in \Fun(\G_\phi)$ with the function
$f$ on $\G$ given by $f(g) = F(g_\phi)$, allowing us to write
\[ \Fun(\G_\phi) \subseteq \Fun_0(\G) .\]
Thus a
generalized measure $\mu$ on $\G$ restricts to a family $\{\mu_\phi\}$
of elements of $\Fun(\G_\phi)^\ast$.   (When $G$ is compact, elements of
$\Fun(\G_\phi)^\ast$ are the same as finite regular Borel measures on
$\G_\phi$.)   We define consistency and uniform
boundedness of such families as in the case of $\A$.

\begin{theorem} \label{thm3} \DOT
Suppose $\mu$ is a generalized measure on $\G$.  For any embedded graph
$\phi$ in $M$, let $\mu_\phi$ denote the restriction of $\mu$ to
$\Fun(\G_\phi)$.  Then $\{\mu_\phi\}$ is a consistent and uniformly
bounded family.  Conversely, if $\{\mu_\phi\}$ is a consistent and
uniformly bounded family, there is a unique generalized measure $\mu$ on
$\G$ whose restriction to $\Fun(\G_\phi)$ is $\mu_\phi$.
\end{theorem}

Proof - The proof follows that of Theorem \ref{thm1}.\qed

We conclude with an alternate description of $\Fun(\A/\G)$.
Recall that functions in $\Fun(\A/\G)$ may be regarded as limits of
$\G$-invariant cylinder functions on $\A$.  At least
when $G$ is amenable (for example, compact, abelian, or an extension of
an amenable group by an amenable group), these are
precisely the same as $\G$-invariant functions on $\A$ that are limits
of cylinder functions:

\begin{theorem} \DOT Suppose that $G$ is amenable.
Then $\Fun(\A/\G)$ is equal to the subalgebra of $\G$-invariant functions
in $\Fun(\A)$.  \end{theorem}

Proof - It is immediate that elements of $\Fun(\A/\G)$ are
$\G$-invariant and lie in $\Fun(\A)$.
To prove the opposite inclusion, suppose $f \in \Fun(\A)$ is
$\G$-invariant.  Then there exists a sequence $f_i \in \Fun_0(\A)$ with
$f_i \to f$.  To show $f \in \Fun(\A/\G)$ it suffices to show the
existence of a sequence of $\G$-invariant elements of $\Fun_0(\A)$
converging to $f$.  We may suppose $f_i \in \Fun(\A_{\phi_i})$.  The
group $\G_{\phi_i}$, being isomorphic to a product of copies of $\G$, is
amenable.  Let $M_i \maps \Fun(\A) \to \Fun(\A)$ denote the result of
averaging over the action of $\G_{\phi_i}$ with respect to an invariant
mean.  Noting that $M_i \maps \Fun(\A_{\phi_i}) \to \Fun(\A_{\phi_i})$,
that $M_i$ is a contraction, and that $M_i f = f$,
 we conclude that $M_i f_i$ is a
sequence of $\G$-invariant elements of $\Fun_0(\A)$ converging to $f$.
\qed

It is not clear whether the hypothesis of amenability is necessary.

\section{Examples}

Now we consider the case where $G$ is compact.  In this case we
construct a generalized measure on $\A$ that we call the uniform
generalized measure.  This generalized measure is invariant under all of
$\Aut(P)$.  It thus projects down to a generalized measure on $\A/\G$
that is $\Diff(M)$-invariant, as described at the end of Section
\ref{prelim}.  The result is the generalized measure on $\A/\G$
constructed for $G = SU(n)$ by Ashtekar and Lewandowski \cite{AL}.  We
also construct a generalized measure on $\G$ called generalized Haar
measure, which is both left- and right-invariant as well as
$\Aut(P)$-invariant.  We also show how to average any generalized
measure on $\A$ against generalized Haar measure on $\G$ to obtain a
$\G$-invariant generalized measure on $\A$.

As in the previous section, we assume $P \mapsto M$ is a $G$-principal
bundle and work in the ``hybrid case,'' case 2 of Section \ref{prelim}.
Let $m$ denote normalized Haar measure on $G$, which is assumed compact.
Suppose $\gamma$ is any path in $M$.  Fixing a trivialization of $P$ at
$\gamma(0)$ and $\gamma(1)$ we obtain an identification of $\A_\gamma$
with a closed and open subset $X \subseteq G$.   Define the
 measure $\mu_\gamma$ on $\A_\gamma$ to be the restriction of $m$
to $\A_\gamma$.  A change of trivialization of $P$ at $\gamma(0)$
changes $X \subseteq G$
by a left translation,
while a change of trivialization at $\gamma(1)$ changes $X$
by a right translation.  Since $m$ is left- and
right-invariant, it follows that $\mu_\gamma$ is independent of
the choice of trivializations at $\gamma(0)$ and $\gamma(1)$.
Recall that for any embedded graph $\phi$
\[         \A_\phi = \prod_{\gamma \in E(\phi)} \A_{\gamma}. \]
Define $\mu_\phi$ to be the finite regular Borel measure on $\A_\phi$
given by the product of the measures $\mu_\gamma$.  Note that
\[         \| \mu_\phi \| \le  1 \]
for all $\phi$.  One can show that $\{\mu_\phi\}$
is a consistent family in the
sense of Theorem \ref{thm1} (for details, see
\cite{AL,B}).
Theorem 1 thus implies the existence of a unique
generalized measure $\mu$ on $\A$ such that for all $\phi$, $\mu_\phi$
is the restriction of $\mu$ to $\Fun(\A_\phi)$.   We call this
generalized measure $\mu_u$ the {\it uniform} generalized measure on
$\A$.   It is easily seen from the natural way in which it was
constructed that $\mu_u$ is $\Aut(P)$-invariant, and thus projects down
to a $\Diff(M)$-invariant generalized measure $\mu_{\rm AL}$ on $\A/\G$.
We call $\mu_{\rm AL}$ the {\it Ashtekar-Lewandowski} generalized
measure on $\A/\G$.

Similarly, given any point $x \in M$, a trivialization of $P$ at $x$
gives an identification of $\G_x$ with $G$.  Using this
identification the Haar measure $m$ on $G$ gives rise
to a measure $\mu_x$ on $\G_x$.  Since $m$ is left- and
right-invariant, $\mu_x$ independent of the choice of trivialization of
$P$ at $x$.  Define the finite regular Borel measure $\mu_\phi$ on
\[       \G_\phi = \prod_{x \in V(\phi)} \G_x \]
to be the product of the measures $\mu_x$.  The family $\{\mu_\phi\}$ is
consistent and uniformly bounded, so by Theorem \ref{thm3} there is a
unique generalized measure $\mu_{\rm H}$ on $\G$ such that for all
$\phi$, $\mu_\phi$ is the restriction of $\mu_{\rm H}$ to $\Fun(\G_\phi)$.
We call this generalized measure {\it generalized Haar measure} on $\G$.
By naturality, $\mu_{\rm H}$ is invariant under the action of $\Aut(P)$
on $\G$.  By the invariance properties of Haar measure, $\mu_{\rm H}$ is
also left- and right-invariant.  That is, the left and right actions of
$\G$ on itself give rise to actions of $\G$ on $\Fun(\G)$, hence dually
on $\Fun(\G)^\ast$, and $\mu_{\rm H}$ is preserved by these actions.

We can convolve generalized measures on $\G$ as follows.
For each  embedded graph $\phi$ one can convolve
finite regular Borel measures on the compact Lie group $\G_\phi$:
\[    \ast \maps   \Fun(\G_\phi)^\ast \times \Fun(\G_\phi)^\ast \to
 \Fun(\G_\phi)^\ast  \]
by the usual formula
\[      (\mu \ast \nu)(f) = \int_{\G_\phi \times \G_\phi} f(gh)\, d\mu(g)
d\nu(h), \]
and one has the bound
 \[         \|\mu \ast \nu\| \le \|\mu \| \, \|\nu\| .\]
If $\phi \hookrightarrow \psi$ there is a natural group homomorphism
$\G_\psi \to \G_\phi$, hence a homomorphism of convolution algebras
$\Fun(\G_\psi)^\ast \to \Fun(\G_\phi)^\ast$.  This allows us to give
$\Fun_0(\G)^\ast$, which is the inverse limit of the spaces
$\Fun(\G_\psi)^\ast$, the structure of an algebra in a unique way such
that all the maps
\[         \Fun_0(\G)^\ast \to \Fun(\G_\phi)^\ast \]
are algebra homomorphisms.  We write the product in $\Fun_0(\G)$ as
$\ast$.   Since this product
satisfies the bound
\[         \|\mu \ast \nu\| \le \|\mu \| \, \|\nu\| ,\]
it extends uniquely by continuity
to a product on $\Fun(\G)^\ast$, again written $\ast$ and
called the {\it convolution} of generalized measures on $\G$.

Moreover, we can average (or convolve) generalized measures on $\A$ against
generalized measures on $\G$ as follows.  For embedded graph $\phi$
the convolution algebra $\Fun(\G_\phi)^\ast$ acts on
$\Fun(\A_\phi)^\ast$,
\[    \ast \maps   \Fun(\G_\phi)^\ast \times \Fun(\A_\phi)^\ast \to
 \Fun(\A_\phi)^\ast  \]
by the usual formula
\[      (\mu \ast \nu)(f) = \int_{\G_\phi \times \A_\phi} f(gA)\, d\mu(g)
d\nu(A), \]
and one has the bound
 \[         \|\mu \ast \nu\| \le \|\mu \| \, \|\nu\| .\]
Using these facts, an inverse limit argument like the one above shows
that the convolution algebra $\Fun(\G)^\ast$ acts on
$\Fun(\A)^\ast$.  Below, we apply this to construct
$\G$-invariant generalized measures on $\A$
from generalized measures on $\A$ by convolution against
Haar generalized measure on $\G$:

\begin{theorem} \label{conv} \DOT Let $\nu$ be a generalized measure on $\A$
and let $\mu_{\rm H}$ denote Haar generalized measure on $\G$.  Then
$\mu_{\rm H} \ast \nu$ is a $\G$-invariant generalized measure on $\A$.
\end{theorem}

Proof - Suppose $f \in \Fun_0(\A)$.  Then $f \in \Fun(\A_\phi)$
for some embedded graph $\phi$, so $f$ depends on $A$ only through
$A_\phi$; we write $f(A) = F(A_\phi)$.  Let $\nu_\phi$
be the restriction of $\nu$ to
$\Fun(\A_\phi)$.  Writing $\mu_\phi$ for Haar measure on $\G_\phi$, we have,
for any $g \in \G$,
\ban
(\mu_{\rm H} \ast \nu)(gf) &=&   (\mu_\phi \ast \nu_\phi)(g_\phi F)   \\
 &=&     \int_{\G_\phi \times \A_\phi} F(g_\phi^{-1} hA_\phi) \,
d\mu_\phi(h) d\nu_\phi(A_\phi)   \\
&=&    \int_{\G_\phi \times \A_\phi} F(h A_\phi) \,
d\mu_\phi(h) d\nu_\phi(A_\phi)   \\
&=&  (\mu_\phi \ast \nu_\phi)(F) \\
&=&   (\mu_{\rm H} \ast \nu)(f)  .\ean
Since $\Fun_0(\A)$ is dense in $\Fun(\A)$, we conclude that
$ ( \mu_{\rm H} \ast \nu)(gf) = (\mu_{\rm H} \ast \nu)(f)$ for all $f \in
\Fun(\A)$, so $\mu_{\rm H} \ast \nu$ is $\G$-invariant.  \qed

\section{Conclusions}

There is much more one can do to generalize
the theory of Lie groups and homogeneous spaces to groups of gauge
transformations and spaces of connections,
using the framework introduced here.  For
example, there is a Hilbert space completion $L^2(\A,\mu_u)$ of
$\Fun(\A)$, on which
$\Aut(P)$ has a unitary representation.  Similarly,
there is a Hilbert space $L^2(\A/\G, \mu_{\rm AL})$ on which $\Diff(M)$
has a unitary representation, and a Hilbert space
$L^2(\G,\mu_{\rm H})$ which is a unitary representation of $\G$
acting by left (or right) translation, as well as a
unitary representation of $\Aut(P)$.  It is still unclear how useful
these structures will be in physics, but they have many of the
properties one would naively expect of nonrigorous constructions using
 ``Lebesgue measure'' on $\A$ and
``Haar measure'' on $\G$.

\vfill
\end{document}